\documentclass[journal=nalefd,manuscript=communication]{achemso}

\usepackage[version=3]{mhchem} % Formula subscripts using \ce{}
\usepackage{graphicx}
\author{Tao Hu}
\affiliation{State Key Laboratory of Surface Physics, Key Laboratory of Micro and Nano Photonic Structures (Ministry of Education), Department of Physics, Fudan University,
Shanghai 200433, China and Collaborative Innovation Center of Advanced Microstructures, Nanjing University, Nanjing Jiangsu 210093, China}

\author{Yafeng Wang}
\affiliation{State Key Laboratory of Surface Physics, Key Laboratory of Micro and Nano Photonic Structures (Ministry of Education), Department of Physics, Fudan University,
Shanghai 200433, China and Collaborative Innovation Center of Advanced Microstructures, Nanjing University, Nanjing Jiangsu 210093, China}

\author{Lin Wu}
\affiliation{State Key Laboratory of Surface Physics, Key Laboratory of Micro and Nano Photonic Structures (Ministry of Education), Department of Physics, Fudan University,
Shanghai 200433, China and Collaborative Innovation Center of Advanced Microstructures, Nanjing University, Nanjing Jiangsu 210093, China}

\author{Long Zhang}
\affiliation{State Key Laboratory of Surface Physics, Key Laboratory of Micro and Nano Photonic Structures (Ministry of Education), Department of Physics, Fudan University,
Shanghai 200433, China and Collaborative Innovation Center of Advanced Microstructures, Nanjing University, Nanjing Jiangsu 210093, China}

\author{Yuwei Shan}
\affiliation{State Key Laboratory of Surface Physics, Key Laboratory of Micro and Nano Photonic Structures (Ministry of Education), Department of Physics, Fudan University,
Shanghai 200433, China and Collaborative Innovation Center of Advanced Microstructures, Nanjing University, Nanjing Jiangsu 210093, China}

\author{Jian Lu}
\affiliation{State Key Laboratory of Surface Physics, Key Laboratory of Micro and Nano Photonic Structures (Ministry of Education), Department of Physics, Fudan University,
Shanghai 200433, China and Collaborative Innovation Center of Advanced Microstructures, Nanjing University, Nanjing Jiangsu 210093, China}

\author{Jun Wang}
\affiliation{State Key Laboratory of Surface Physics, Key Laboratory of Micro and Nano Photonic Structures (Ministry of Education), Department of Physics, Fudan University,
Shanghai 200433, China and Collaborative Innovation Center of Advanced Microstructures, Nanjing University, Nanjing Jiangsu 210093, China}

\author{Song Luo}
\affiliation{State Key Laboratory of Surface Physics, Key Laboratory of Micro and Nano Photonic Structures (Ministry of Education), Department of Physics, Fudan University,
Shanghai 200433, China and Collaborative Innovation Center of Advanced Microstructures, Nanjing University, Nanjing Jiangsu 210093, China}

\author{Zhe Zhang}
\affiliation{State Key Laboratory of Surface Physics, Key Laboratory of Micro and Nano Photonic Structures (Ministry of Education), Department of Physics, Fudan University,
Shanghai 200433, China and Collaborative Innovation Center of Advanced Microstructures, Nanjing University, Nanjing Jiangsu 210093, China}

\author{Liming Liao}
\affiliation{State Key Laboratory of Surface Physics, Key Laboratory of Micro and Nano Photonic Structures (Ministry of Education), Department of Physics, Fudan University,
Shanghai 200433, China and Collaborative Innovation Center of Advanced Microstructures, Nanjing University, Nanjing Jiangsu 210093, China}

\author{Shiwei Wu}
\affiliation{State Key Laboratory of Surface Physics, Key Laboratory of Micro and Nano Photonic Structures (Ministry of Education), Department of Physics, Fudan University,
Shanghai 200433, China and Collaborative Innovation Center of Advanced Microstructures, Nanjing University, Nanjing Jiangsu 210093, China}

\author{S. C. Shen}
\affiliation{State Key Laboratory of Surface Physics, Key Laboratory of Micro and Nano Photonic Structures (Ministry of Education), Department of Physics, Fudan University,
Shanghai 200433, China and Collaborative Innovation Center of Advanced Microstructures, Nanjing University, Nanjing Jiangsu 210093, China}

\author{Zhanghai Chen}
\email{zhanghai@fudan.edu.cn}
\affiliation{State Key Laboratory of Surface Physics, Key Laboratory of Micro and Nano Photonic Structures (Ministry of Education), Department of Physics, Fudan University,
Shanghai 200433, China and Collaborative Innovation Center of Advanced Microstructures, Nanjing University, Nanjing Jiangsu 210093, China}

\title[An \textsf{achemso} demo]
  {Strong coupling between Tamm plasmon polariton and two dimensional semiconductor excitons}

\abbreviations{IR,NMR,UV}

\begin{document}
%%%%%%%%%%%%%%%%%%%%%%%%%%%%%%%%%%%%%%%%%%%%%%%%%%%%%%%%%%%%%%%%%%%%%
%% The manuscript does not need to include \maketitle, which is
%% executed automatically.  The document should begin with an
%% abstract, if appropriate.  If one is given and should not be, the
%% contents will be gobbled.
%%%%%%%%%%%%%%%%%%%%%%%%%%%%%%%%%%%%%%%%%%%%%%%%%%%%%%%%%%%%%%%%%%%%%
\begin{abstract}
Two dimensional (2D) semiconductor materials of transition-metal dichalcogenides (TMDCs) manifest many peculiar physical phenomena in the light-matter interaction. Due to their ultrathin property, strong interaction with light and the robust excitons at room temperature, they provide a perfect platform for studying the physics of strong coupling in low dimension and at room temperature. Here we report the strong coupling between 2D semiconductor excitons and Tamm plasmon polaritons (TPPs). We observe a Rabi splitting of about 54 meV at room temperature by measuring the angle resolved differential reflectivity spectra and simulate the theoretical results by using the transfer matrix method. Our results will promote the realization of the TPP based ultrathin polariton devices at room temperature.
\end{abstract}

\section{Introduction}

Strong coupling between photons and excitons, so called "exciton polariton", is a fascinating topic in solid state physics and has attracted much attention in recent years. Numerous novel phenomena have been observed, e.g., Bose Einstein Condensation of exciton polariton,\cite{kasprzak1,balili2,deng3} superfluidity \cite{amo4}, quantum vortices \cite{nardin5}, entangled photon pairs \cite{johne6} and polariton bistability,\cite{ballarini7,espinosa8} etc. These exciting researches have also promoted the development of novel devices such as electrically injected polariton light emitting diode, \cite{tsintzos9,sapienza10,schneider10} low threshold polariton laser, \cite{bhattacharya11,bhattacharya12} spin-optronic devices, \cite{shelykh13} optical switching \cite{piccione14} and polaritonic logic circuits,\cite{menon15} etc.

In order to obtain well confined photons and thus observe the exciton-polariton effect, various optical microcavities, such as distributed Bragg reflector (DBR) based planar microcavity, \cite{stanley16} tunable open cavity, \cite{schwarz17} two dimensional photonic crystal \cite{akahane18} and microdisk cavity with whispering gallery mode, \cite{srinivasan19} were designed. Recently, it is shown that one can also confine the light even without a cavity, e.g., a surface wave will form at the interface between a metal and a DBR. This is the so-called TPPs. \cite{kavokin20,kaliteevski21,gazzano22,bruckner23} In contrast to conventional surface plasmons, TPPs can be excited by direct optical excitation as their dispersion lies in the light cone given by $k=\omega/c$ where $\omega$ is the angular frequency and $k$ is the in-plane component of the wave vector of light, and TPPs have both TE and TM polarization. Moreover, since in the strong light-matter coupling regime Rabi splitting is proportional to the amplitude of the vacuum field, one can increase it through decreasing the mode volume. Compared with traditional DBR-DBR cavity, TPP mode gives smaller mode volume due to its surface wave nature.

Exciton polariton effect has been studied in various semiconductor microcavity systems, including ZnO \cite{sun24,xie25} and GaN \cite{das26} with large exciton binding energy which can work at room temperature. Meanwhile the organic semiconductors also exhibit great prospect due to their strong exciton binding energies, high oscillator strengths and high quantum yields. \cite{kena27,lagoudakis28} Very recently, a new candidate for polaritonics, i.e., TMDCs, has attracted much attention\cite{liu29,dufferwiel30,liu31,lundt32,liu33} due to their distinct electronic, mechanical, thermal and optical properties when it is thinned to monolayer.\cite{xia34} These materials change from indirect to direct bandgap with the transition from bulk to monolayer and the coupled spin and valley physics in monolayer TMDCs materials leads to the valley Hall effect.\cite{mak35} Meanwhile, their robust excitons at room temperature thanks to their large exciton binding energy (~0.5 - 1.0 eV ) and the excellent optical qualities make them of great potential in the physics of strong coupling between light and matter at room temperature. Moreover, as the TPP structure do no contain the cavity layer, it is easier to fabricate compared to the traditional microcavity systems. Furthermore, due to the surface wave nature of the TPP mode, it is highly advantageous for the strong coupling with the 2D semiconductor materials. However, the strong coupling between the TPP mode and the monolayer TMDCs materials is yet to be demonstrated.

In this paper, we demonstrate the strong coupling between TPPs and the A excitons in monolayer MoS$_{2}$ and observe a Rabi splitting of about 54 meV at room temperature by the angle resolved reflectivity spectroscopy. Theoretical simulation by using the transfer matrix method \cite{kavokin20,kaliteevski21,kavokin39} agrees well with the experimental results.

\section{Results and discussion}

\begin{figure}[htb]
     \includegraphics[width=13cm]{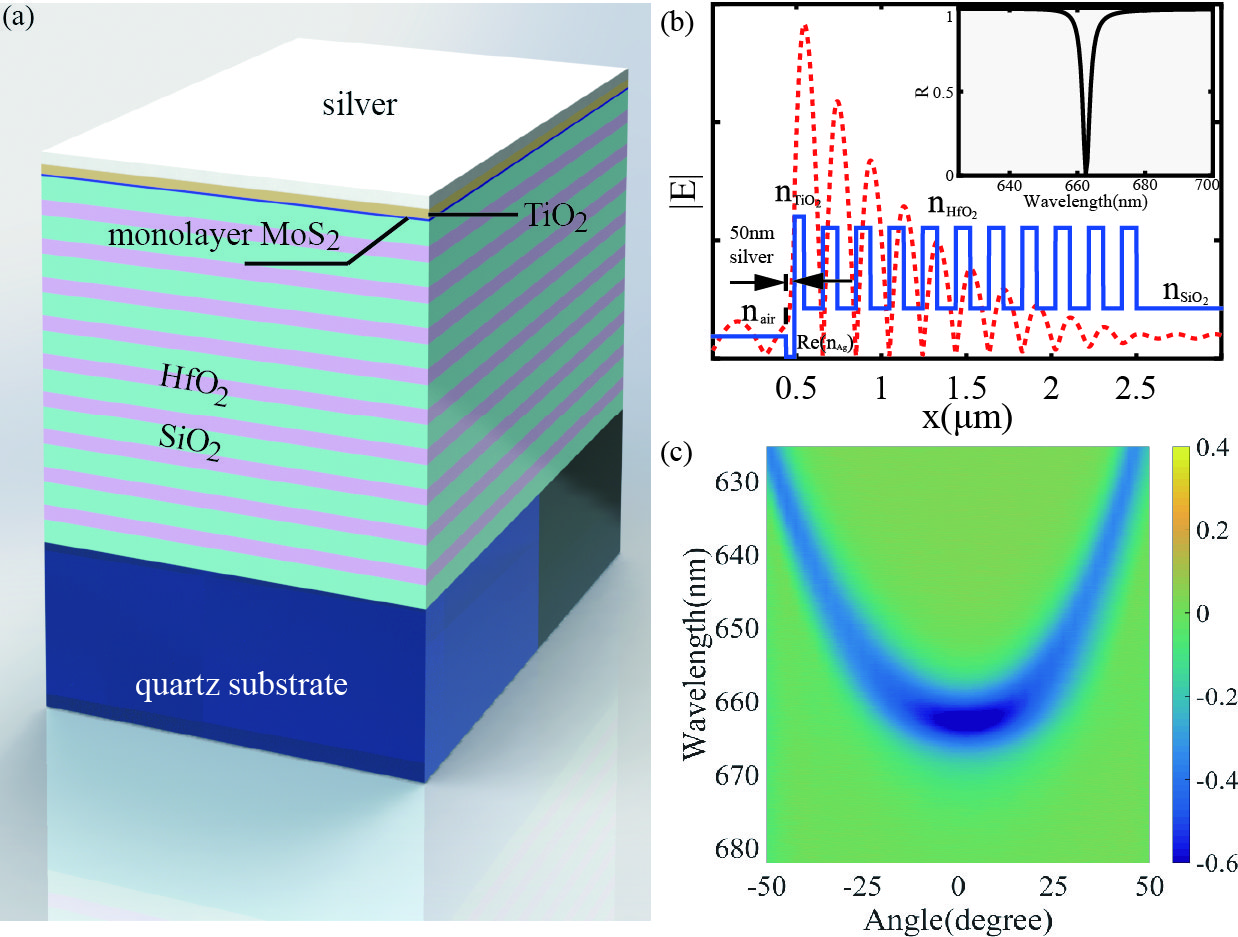}
     \caption{(a) Schematic of the sample structure. (b) (normal incidence) Theoretically calculated absolute electric field distribution of the TPP mode with wavelength of 663 nm. The inset picture shows the reflectivity spectrum of the structure in (a) without monolayer MoS$_{2}$ embedded in it. (c) Experimentally detected angle resolved differential reflectivity spectra of the TPP mode.}
\end{figure}

Figure 1a shows the sample structure schematically. 10 pairs of HfO$_{2}$/SiO$_{2}$ DBR on the quartz substrate are grown by electron beam evaporation. The monolayer MoS$_{2}$ is exfoliated mechanically from bulk and then transferred onto the DBR by using polydimethylsiloxane (PDMS).\cite{choi37} To form the TPP state, the top layer of TiO$_{2}$ of 56 nm and the silver film of 50 nm are deposited in the following electron beam evaporation process. In order to obtain the maximum coupling between TPP modes and the monolayer MoS$_{2}$, we need to position the MoS$_{2}$ layer in the region where the strongest amplitude of electric field is presented as shown in figure 1b. Via using the transfer matrix method we give the theoretical calculation of the TE polarized TPP mode and the distribution of absolute electric field at normal incidence in figure 1b. The parameters used in the calculation are $n_{SiO_{2}}=1.5, n_{HfO_{2}}=1.95, n_{TiO_{2}}=2.12$ and $n_{SiO_{2}}l_{SiO_{2}}=n_{HfO_{2}}l_{HfO_{2}}=$$\pi$$c/2\omega_{0}$, where $\omega_{0}$ corresponds to the Bragg frequency with $\hbar$$\omega_{0}$=1.87 eV. The relative permittivity of silver is described by the Drude model:
\begin{eqnarray}
\varepsilon_{Ag}(\omega)&=\varepsilon_\infty-\frac{\omega_{p}^2}{\omega^2+i\gamma\omega},n_{Ag}&=\sqrt{\varepsilon_{Ag}(\omega)}
\end{eqnarray}
where $\varepsilon_{\infty}=5, \hbar\omega_{p}=9$ eV,$\hbar\gamma=18$ meV. We confirm the TPP state in the structure by measuring angle resolved differential reflectivity spectra $\bigtriangleup$$R/R_{0}=(R-R_{0})/R_{0}$ (shown in figure 1c), where $R$ is the reflectivity of the TPP sample and $R_{0}$ is that of the silver reflector. The numerical aperture (NA) of the objective lens we use is 0.75 which allows us to detect the angular range of $\pm$48.6$^\circ$. The broadband light source is a tungsten halogen lamp and the angle resolved differential reflectivity spectroscopic setup is similar to that described in Ref [32]. From figure 1c, one can see a dip at 663 nm with half width at half maximum (HWHM) of $\hbar\Gamma_{TPP}$=14 meV. This implys a quality factor of the TPP mode of 130 at normal incidence. A parabolic dispersion of the TPP mode with the angle of the incidence light varying from 0$^\circ$ to 48.6$^\circ$ is observed.

\begin{figure}[!t]
     \includegraphics[width=13cm]{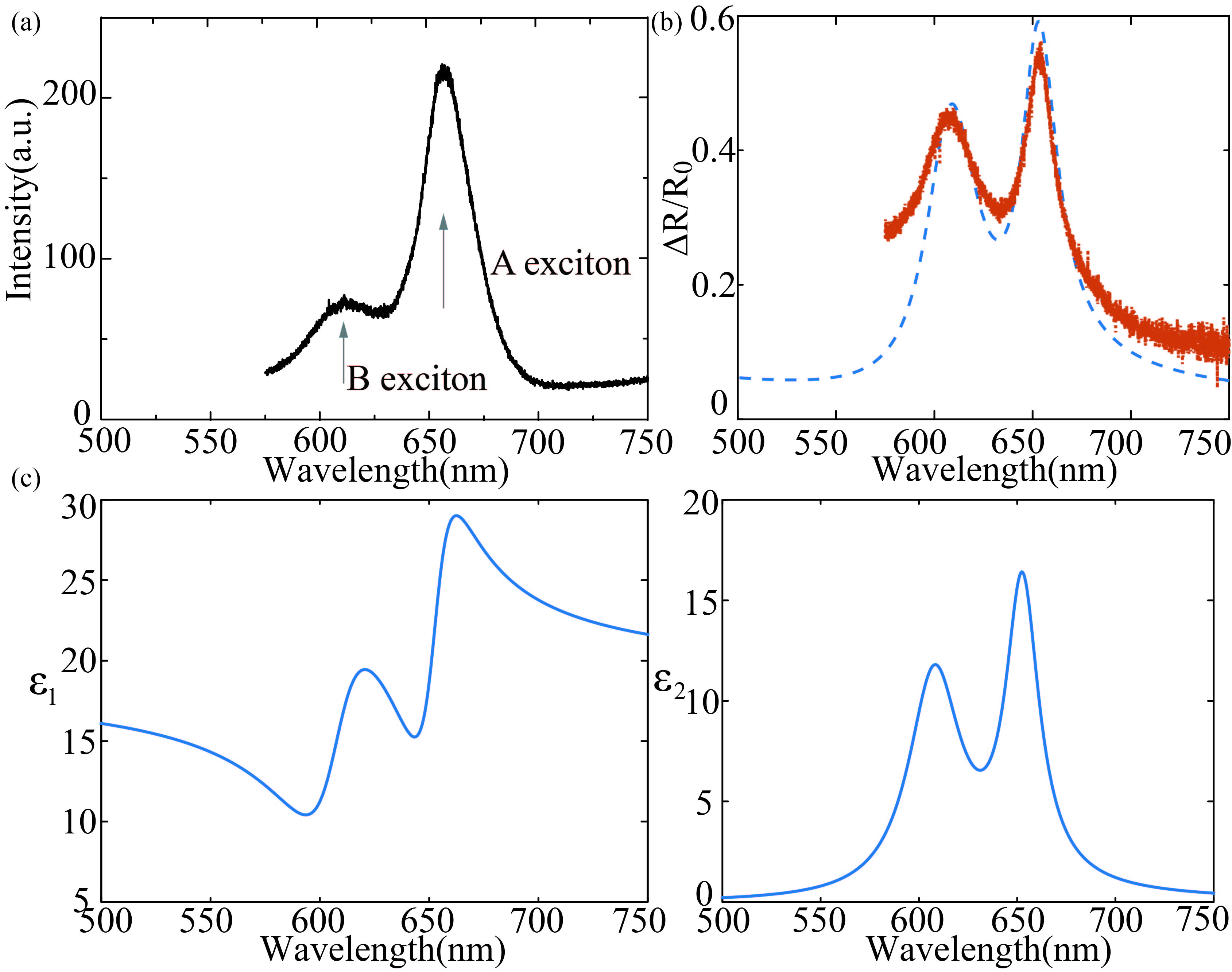}
     \caption{(a) Photoluminescence of the monolayer MoS$_{2}$. The gray arrows indicate the A and B excitons in monolayer MoS$_{2}$. (b) The experimental (orange dotted line) and the theoretically simulated results (blue dashed line) of the the differential reflectivity spectrum of the monolayer MoS$_{2}$ on the PDMS substrate. (c) The real and imaginary parts of the complex dielectric function of the monolayer MoS$_{2}$.}
\end{figure}

The optical qualities of monolayer MoS$_{2}$ are examined by measuring micro-photoluminescence (PL) spectroscopy and differential reflectivity spectroscopy, as shown in figure 2a and 2b. In the PL experiments, the excitation is a 532 nm continuous wave laser line. From figure 2a, one can see two pronounced PL peaks at 652.5 nm and 610 nm associated with the A and B excitons, respectively\cite{mak38} and the dominant peak is 652.5 nm arising from the transition of the A excitons. The PL spectrum indicates the excellent optical quality of the monolayer MoS$_{2}$ and demonstrates that the thickness of exfoliated MoS$_{2}$ is single layer. In the measurement of the reflectance of our samples, the differential reflectivity is defined as $\bigtriangleup$$R/R_{0}=(R_{sample}-R_{0})/R_{0}$, where $R$$_{sample}$ is the reflectivity of the monolayer MoS$_{2}$ on the substrate (PDMS) and $R$$_{0}$ is that of the substrate. Two clear absorption peaks are observed which are consistent with the two peaks in figure 2a. Moreover, from the reflectance measurements of the samples, we can deduced the complex dielectric function of the monolayer MoS$_{2}$. The dielectric function of the monolayer MoS$_{2}$ is modelled by the multi-Lorentzian oscillators:\cite{li40}
\begin{equation}
\varepsilon(\omega)=\varepsilon_{b}+\Sigma_{i=A,B}\frac{f_{i}}{\omega_{i}^{2}-\omega^2-i\Gamma_{i}\omega}=\varepsilon_{1}+i\varepsilon_{2}
\end{equation}
where $\varepsilon_{b}$ is the background dielectric function , $f$ is the oscillator strength and the parameter i=A, B correspond to the A and B excitons in MoS$_{2}$. $\hbar\omega_{A}$= 1.9 eV, $\hbar\omega_{B}=2.04$ eV, $\hbar\Gamma_{A}$=60 meV and $\hbar\Gamma_{B}$=110 meV are fitted from the PL spectrum in figure 2a. By the simulation of the differential reflectivity spectrum (blue dashed line) which is in good agreement with the experimental result (orange dotted line), we can obtain the other parameters which are $\varepsilon_{b}=18,\hbar^{2}f_{A}=1.7$ ev$^{2}$, $\hbar^{2}f_{B}=$2.5 eV$^{2}$. In the simulation the refractive index of PDMS is set to be 1.5. The derived real and imaginary part of the complex dielectric function of the monolayer MoS$_{2}$ are given in figure 2c which are consistent with the reported experimental results.\cite{li40}

\begin{figure}[!t]
     \includegraphics[width=13cm]{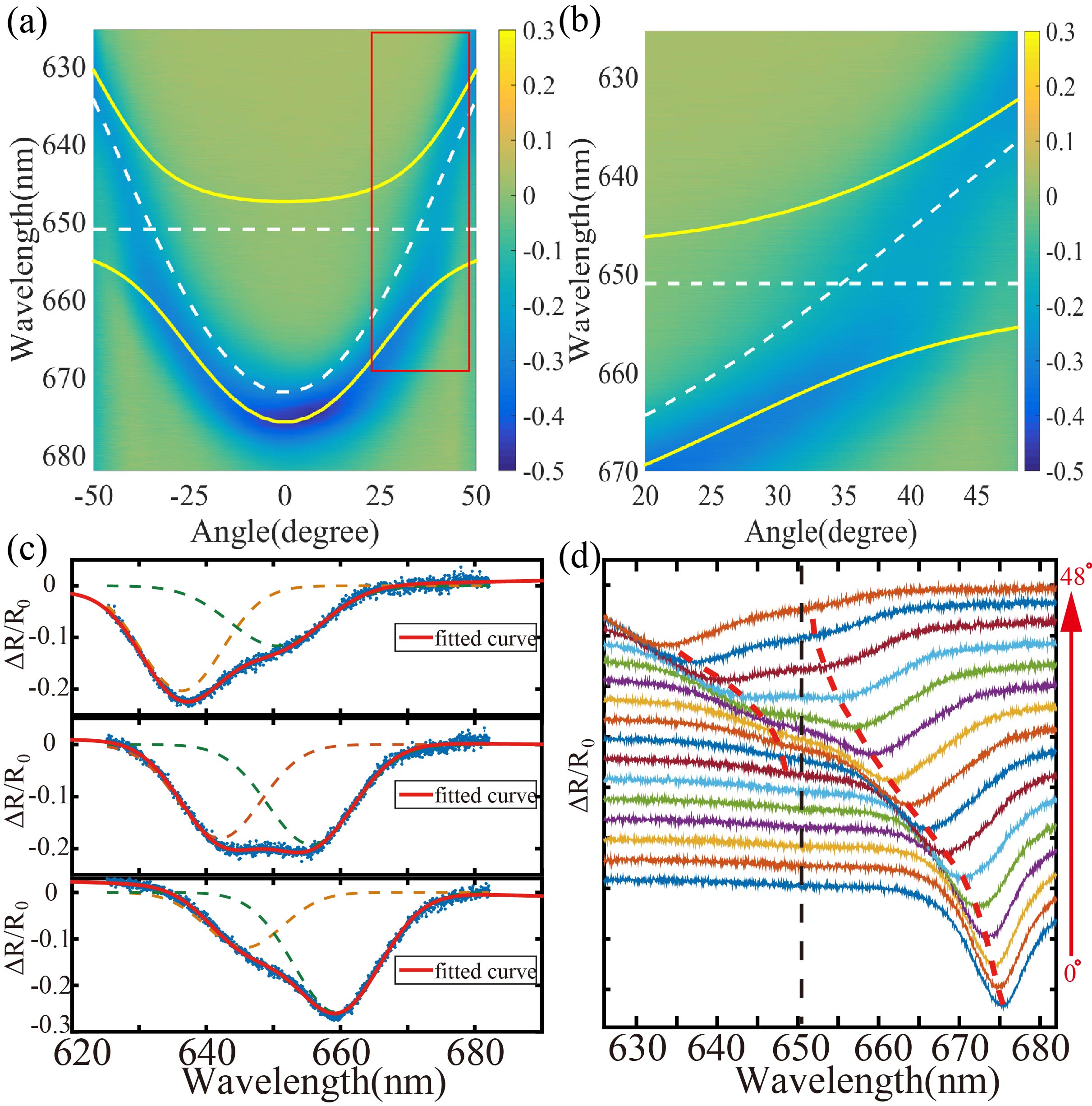}
     \caption{(a) Angle resolved differential reflectivity spectra of the TPP sample with the monolayer MoS$_{2}$ embedded in it. The white dashed line indicates the A exciton energy of MoS$_{2}$ and the dispersion of the TPP mode. The yellow lines represent the dispersion of the LPB and UPB. (b) The enlarged color map of the red rectangle in (a). (c). The fitted curves (red line) of the spectra extracted from (a) with angle of 33$^{\circ}$, 39$^{\circ}$, 45$^{\circ}$, respectively. The orange dashed lines and green dashed lines are the Gaussian fitting of the UPB and the LPB and the blue dotted lines are the experimental results.(d) Sixteen extracted spectra of (a) with angle varying from 0$^{\circ}$ to 48$^{\circ}$. The black dashed line shows the A exciton energy and the red dashed lines represent the fitted minima of each spectrum.}
\end{figure}

By measuring the angle resolved differential reflectivity spectra, we demonstrate the strong coupling between the TPP modes and A excitons in monolayer MoS$_{2}$. From figure 3a, two pronounced dips which correspond to the upper polariton branch (UPB) and the lower polariton branch (LPB) are seen at the angle of $\pm 38^{\circ}$ when the TPP mode is resonant with the A excitons. The detailed part of the red rectangle in figure 3a is displayed in figure 3b. With the increase of angle from 38$^{\circ}$ the UPB is more TPP like while the LPB is more exciton like and vice versa for the decrease of the angle from 38$^{\circ}$. In figure 3a only UPB is observed in large angle (>38$^{\circ}$) and LPB in small angle (<38$^{\circ}$), this is due to the broad linewidth of the A excitons at room temperature and the low contrast in the mapping of the differential reflectivity spectra. For clarity, we extract the differential reflectivity spectra of figure 3a and fit the dips of these spectra. Three of the fitting results are shown in figure 3c with angle of 33$^{\circ}$, 39$^{\circ}$, 45$^{\circ}$, respectively and two clear dips which are shown by the orange dashed line and green dashed line are seen in each spectrum. The fitting results from 0$^{\circ}$ to 48$^{\circ}$ are presented in figure 3d and the red lines indicate the minima of these spectra. These dips reveal clear anti-crossing feature when they approach the A exciton energy indicated by the black dashed line. We calculate the theoretical polariton dispersion with the coupling oscillator model expressed as: \cite{liu29}
\begin{equation}
\left(
\begin{array}{ccc}
    E_{exciton} & V\\
    V & E_{TPP}(\theta)\\
  \end{array}
\right)=E(\theta)
\left(
\begin{array}{ccc}
a\\
b\\
\end{array}
\right)
\end{equation}
where the E$_{exciton}$ is the energy of the A exciton, and E$_{TPP}(\theta)$ is the energy of the TPP mode which is calculated by using the transfer matrix method. V is the interaction strength between the TPP modes and A excitons in MoS$_{2}$. $E(\theta)$ is the eigen energy of the polariton branch given by:
\begin{equation}
E_{LPB,UPB}(\theta)=\frac{1}{2}\left[E_{exciton}+E_{TPP}(\theta)\pm\sqrt{4V^{2}+(E_{exciton}-E_{TPP}(\theta))^{2}}\right]
\end{equation}
and |a|$^{2}$, |b|$^{2}$ represents the exciton and TPP component of the corresponding polariton branch, respectively. The calculated dispersions are shown by the yellow lines in figure 3a and 3b. The white dashed line indicates the energy of the A excitons and the dispersion of the TPP mode. From the theoretical results, the detuning $\bigtriangleup$$E$=$E_{TPP}-E_{exciton}$= -60 meV and the fitted interaction potential $V$ is about 27 meV indicating that the Rabi splitting $\hbar\Omega$ is 54 meV. To confirm the system is in the strong coupling regime, we extract the HWHM of the A excitons from the PL spectrum in figure 2 with $\hbar\Gamma_{A}$= 60 meV. It is proved that the strong coupling condition for light-matter interaction is satisfied with $\hbar\Omega>(\hbar\Gamma_{A}+\hbar\Gamma_{TPP})/2$. However, the strong coupling between the TPP modes and the B excitons is not observed due to the large detuning between the TPP modes and B excitons.\cite{liu29}

We also simulate the angle resolved differential reflectivity spectra of the strong coupling between the A excitons and TPP modes by using transfer matrix method which are given in figure 4. During the simulation, the thickness of monolayer MoS$_{2}$ is set to be 0.65 nm which is a common thickness for monolayer TMDCs materials and the complex refractive index of monolayer MoS$_{2}$ is derived from equation (2). The simulated results give features similar to the experimental results. The UPB is only observable at large angle while the LPB is observable at small angle from the color map in figure 4a. The extracted spectra in figure 4b show clear anti-crossing feature indicating the formation of the exciton polariton state. The vague in the spectra is mainly due to the broad linewidth of the A excitons at room temperature and the weak interaction strength between the TPP modes and the A excitons in the monolayer MoS$_{2}$ as the overlap between light field and the monolayer MoS$_{2}$ is very small due to the ultrathin property of 2D TMDCs materials. To confirm this, we increase the quality factor of the TPP mode with $\hbar\Gamma_{TPP}$= 1.7 meV and the calculated spectra in figure 5a are nearly the same as figure 4a. That means the decreasing of the linewidth of the TPP modes from 14 meV to 1.7 meV do not increase the visibility of the LPB and UPB. Hence, we simulate the spectra with decreasing $\hbar\Gamma_{A}$ to 30 meV and 10 meV artificially which are shown in figure 5b and 5c with clear LPB and UPB. If we enhance the oscillator strength such as utilizing the multi quantum well structure, the splitting between LPB and UPB can also be seen in the angle resolved differential reflectivity spectra shown in figure 5d. The simulation results of change of detuning are shown in figure 6.

\begin{figure}[!t]
     \includegraphics[width=13cm]{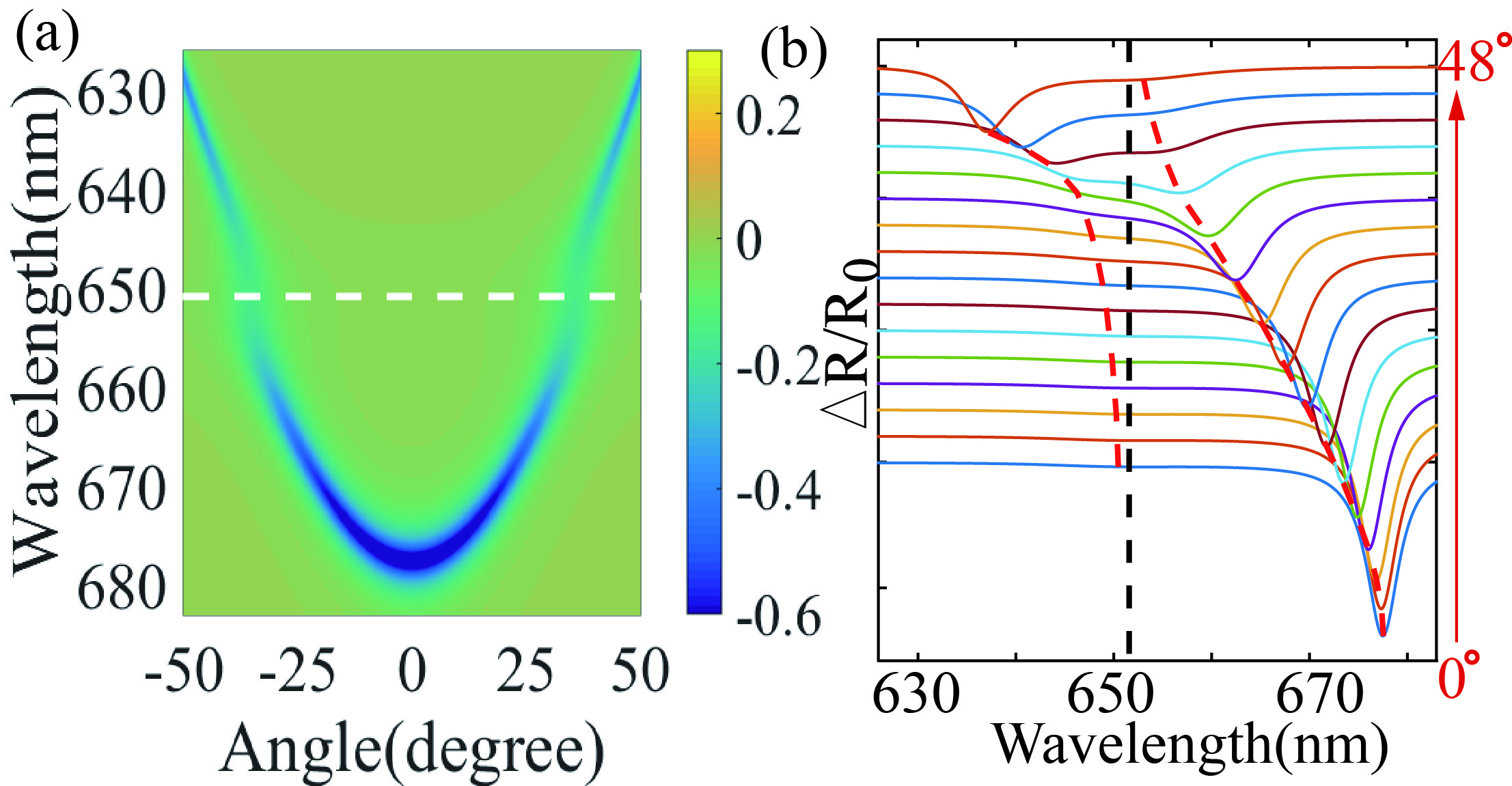}
     \caption{(a) Simulated results of angle resolved differential reflectivity spectra of the structure in figure 1a. The white dashed line indicates the A exciton energy of MoS$_{2}$. (b) Sixteen extracted spectra of (a) with angle varying from 0$^{\circ}$ to 48$^{\circ}$. The black dashed line shows the A exciton energy and the red dashed lines are the fitted minima of each spectrum.}
\end{figure}

The realization of strong coupling between photons and 2D semiconductor materials is of great importance due to the excellent optical qualities, stable excitons at room temperature and the unique nature of the electron state of the 2D semiconductor materials. These materials provide a perfect platform for the investigation of exciton polariton effect in a low dimensional system and manipulation of exciton polariton emission by the valley degree of freedom. These unique properties will lead to the fabrication of novel controllable circular polarized polariton devices. Meanwhile, the electroluminescence of the 2D semiconductor has been reported in different structures\cite{ye42,ross43,cheng44} and the lasing actions\cite{salehzadeh45,ye46,wu47} of such materials are also realized. Hence, the fabrication of the 2D semiconductor materials based polariton laser will be a challenging task in the future.

In summary, we demonstrate the strong coupling between the TPP mode and the A excitons in monolayer MoS$_{2}$ with a Rabi splitting of 54 meV. By measuring angle resolved differential reflectivity spectra we observe the anticrossing feature of the LPB and UPB indicating the formation of the exciton polariton state in the system. By using the transfer matrix method we give the simulation of strong coupling effect which agrees well with the experimental results. The realization of strong coupling between TPP and 2D TMDCs materials paves the way for the fabrication of real polariton devices in the TPP structure and reveal the possibility of electrically injected polariton laser in the future.
\begin{acknowledgement}
The work is funded by the National Science Fund for Distinguished Young Scholars (No. 11225419) and Program of Shanghai Subject Chief Scientist (No. 14XD1400200).
\end{acknowledgement}

\bibliography{tpp}
\begin{suppinfo}
\begin{figure}[htb]
     \includegraphics[width=13cm]{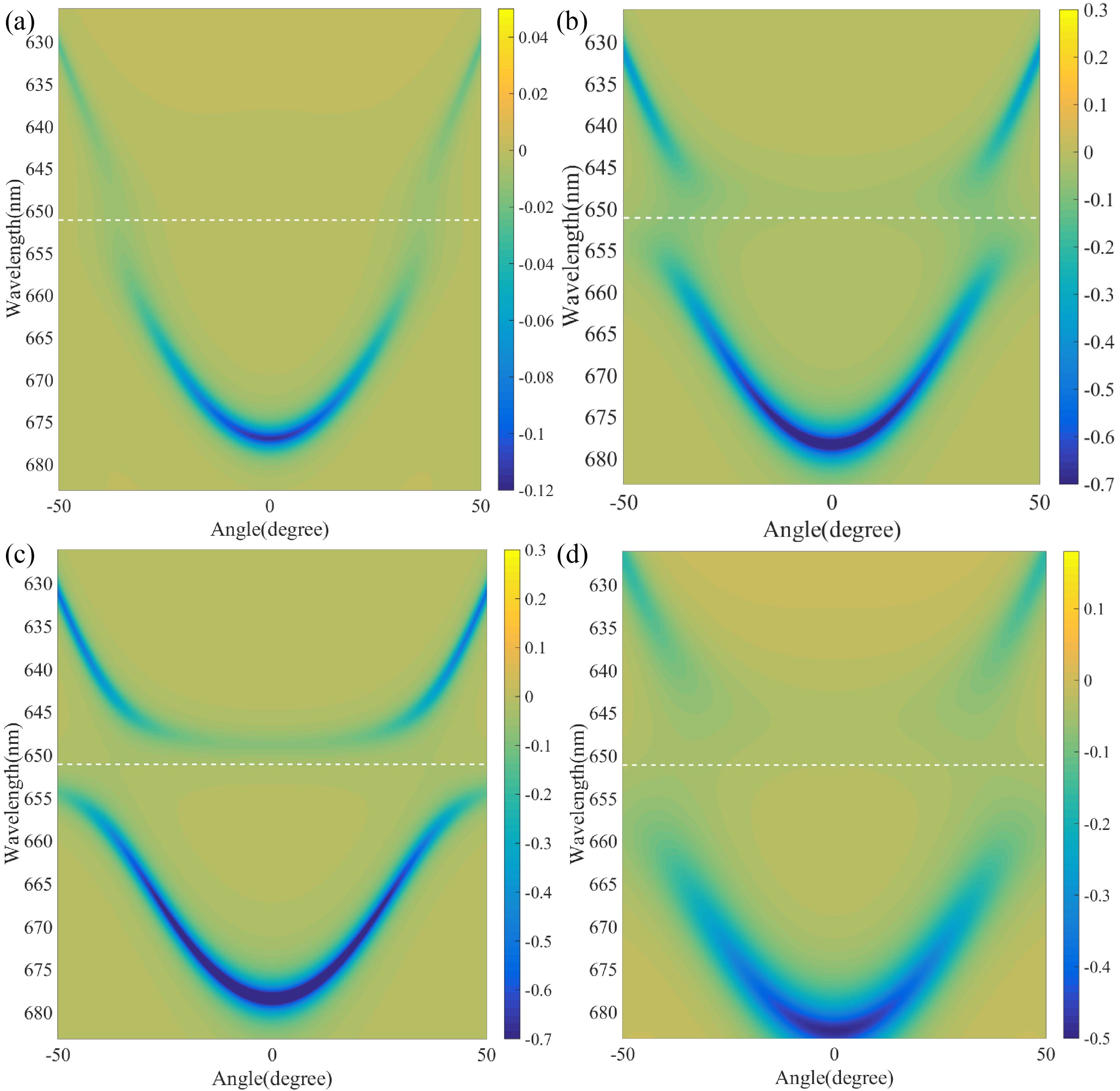}
     \caption{Simulated strong coupling effect with artificially adjusted parameters for the TPP linewidth, exciton linewidth and oscillator strength. The white dashed line corresponds to the A exciton energy in the monolayer MoS$_{2}$. (a)$\hbar\Gamma_{TPP}$=1.7 meV. (b) $\hbar\Gamma_{A}$=30 meV. (c) $\hbar\Gamma_{A}$=10 meV. (d) Oscillator strength $\hbar^{2}f_{A}$=8 eV$^{2}$, which means the calculated Rabi splitting is about 69 meV}
\end{figure}
It is obvious that the increase of the quality factor of the TPP mode can not increase the visibility of the UPB and LPB. But if we put the sample at low temperature to obtain narrower exciton linewidth, such as 30 meV or 10 meV in figure 5b and 5c, we will observe a clear anti-crossing feature of the UPB and LPB. An other way to observe the clear strong coupling effect is to enhance the coupling strength, i.e. $f$, between the TPP mode and the excitons like the multi quantum well structure. Figure 5d is the simulated result for $\hbar^{2}f_{A}$=8 eV$^{2}$ which corresponds to a Rabi splitting of 69 meV, the UPB and LPB are easy to distinguish from each other.
\begin{figure}[htb]
     \includegraphics[width=13cm]{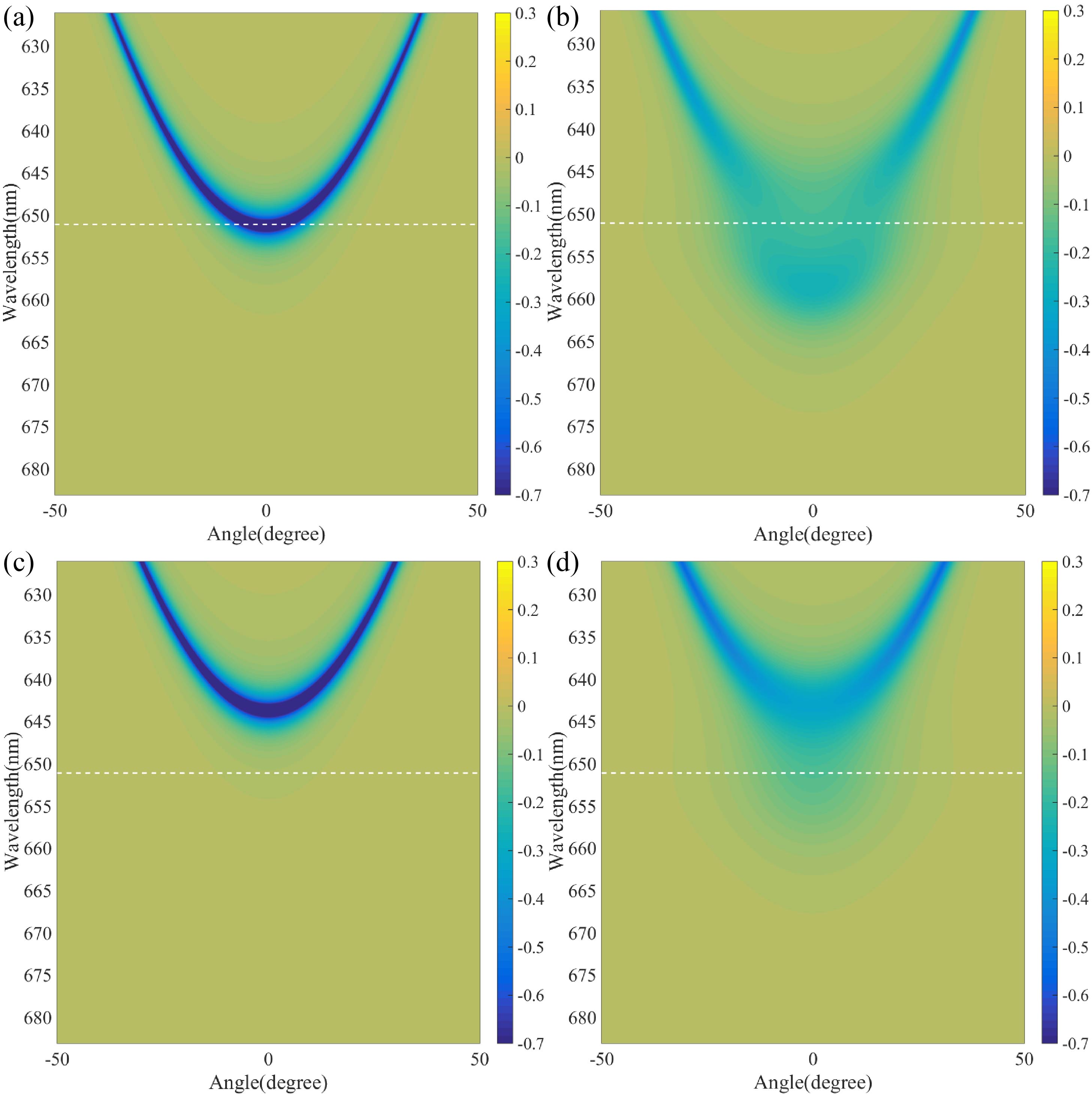}
     \caption{Theoretically simulated strong coupling with different detunings. (a), (c) are the simulated pure TPP dispersions with detuning = 0 meV and 20 meV and (b), (d) are the corresponding strong coupling dispersions. The dashed white line indicates the A exciton energy in the monolayer MoS$_{2}$.}
\end{figure}
\end{suppinfo}


\end{document}